\documentclass[10pt]{article}

\def\etal{{\it et al.}}

\def\th{\theta}
\def\ka{\kappa}
\def\ga{\gamma}
\def\de{\delta}
\def\si{\sigma}
\def\ch{\chi}
\def\om{\omega}
\def\De{\Delta}

\def\ket#1{|{#1}\rangle}
\def\half{{\textstyle{1\over 2}}}
\newcommand{\beq}{\begin{equation}}
\newcommand{\eeq}{\end{equation}}
\newcommand{\bea}{\begin{eqnarray}}
\newcommand{\eea}{\end{eqnarray}}
\newcommand{\rf}[1]{(\ref{#1})}

\begin{document}

\title{Tests of Lorentz symmetry using
antihydrogen\footnote{Proceedings of the
37th Winter Colloquium on the Physics of Quantum Electronics,
held January 2-6, 2007, in Snowbird, Utah, USA.}
}

\author{Neil Russell
\\\vspace{6pt}
Physics Department,\\ Northern Michigan University,\\ Marquette, MI 49855, U.S.A.
}

\date{}

\maketitle

\begin{abstract}
Signals of CPT and Lorentz violation
are possible in the context of
spectroscopy using hydrogen and antihydrogen.
We apply the Standard-Model Extension,
a broad framework for Lorentz breaking in physics,
to various transitions in the hydrogen and antihydrogen spectra.
The results show an unsuppressed effect in
the transition between the upper two hyperfine sublevels
of the ground state of these systems.
We also discuss related tests in Penning traps,
and recent work on Lorentz violation in curved spacetime.
\end{abstract}

\section{Introduction}
The theory of General Relativity and the Standard Model of particle physics
are, by construction, Lorentz-symmetric theories.
Numerous experiments have confirmed
that Lorentz symmetry,
and the closely associated CPT symmetry \cite{greenberg},
are features of nature
at currently accessible precisions.
At extremely high energies near the $10^{19}$~GeV Planck scale,
where the two theories are expected to merge,
it is possible that Lorentz violations may occur.
Such violations are possible, for example,
in string theory with spontaneous symmetry
breaking \cite{theory}.
Although the expected energies are not directly attainable
in experiments,
the possibility exists of detecting suppressed effects
of Planck-scale physics in suitable high-precision experiments.
There exists an effective field theory
providing a full description of the unconventional signals
that may be seen in current experiments,
regardless of their origin.
It consists of the
Standard Model of particle physics coupled to General Relativity,
together with all terms constructed from operators for Lorentz violation.
It is referred to as the Standard-Model Extension,
or SME \cite{SME:minkowski,gr1}.

For the last decade,
the SME has provided a unified framework
allowing the isolation of unsuppressed signals for possible violations
and defining specific coefficients for experimental measurements.
A variety of theoretical issues in the photon sector
have been investigated
\cite{km,em:theor,hariton,radiative_corr}.
A large number of electromagnetic experiments
have been done, including ones with
microwave and optical cavities \cite{photon:cavities},
scattering processes and Cerenkov radiation \cite{cerenkov},
and the Casimir effect \cite{casimir}.
Recent studies of Lorentz violation involving
gamma-ray bursts and the cosmic microwave background
have been completed \cite{grbcmb}.
Theoretical and experimental studies of Lorentz violation in the context of
electron physics
have been done with
torsion pendula \cite{torsion:theor,torsion:exp}
and Penning traps \cite{Penning:theory,Penning:expt}.
Spectroscopy with atomic clocks
has been used to produce stringent bounds
on coefficients for Lorentz violation
\cite{cc,cc:expt,protneut},
and proposals for space-based experiments have been made
\cite{cc:space}.
Theoretical and experimental results have been obtained
for
muons \cite{muons},
neutrinos \cite{neutrino},
the Higgs \cite{higgs},
the early universe \cite{baryogen},
neutral mesons \cite{mesons},
noncommutative geometry \cite{noncomm},
and other systems
\cite{sme:other}.
Further details of this broad and expanding area
can be found in various reviews \cite{SMEreviews}.

The majority of the experiments to date have focused
on the flat spacetime limit of the SME \cite{SME:minkowski}.
Recent studies have shown that the SME framework
extends into the context of
curved spacetime in a consistent manner \cite{gr1}.
In this treatment,
the coefficients for Lorentz violation vary with position,
and the spin effects of matter
are introduced using the vierbein formalism.
A number of questions about Lorentz breaking
in curved spacetime have been answered.
One of these relates to whether the symmetry breaking
is explicit or spontaneous \cite{gr2}.
It has been found that explicit breaking is
incompatible with generic Riemann-Cartan spacetimes,
except perhaps in more general geometries
such as occur in Finsler spaces
\cite{gr1,finsler}.
On the other hand, spontaneous breaking
can be introduced in a consistent manner.
Another question relates to the massless modes
that are expected
in spontaneous symmetry breaking.
There are
10 possible such Nambu-Goldstone modes associated with
the six generators for Lorentz transformations and
the four generators for diffeomorphisms.
The results are consistent with the known massless particles in nature.
Renormalization of quantum electrodynamics in curved spacetime
has also been investigated \cite{renorm}.

The idea of using a potential to spontaneously break Lorentz symmetry,
thus enforcing a nonzero vacuum value for a tensor field,
was introduced by Kosteleck\'y and Samuel \cite{theory}.
Several models for such fields have been created as useful test cases,
and include the bumblebee field \cite{gr2} and the cardinal field \cite{gr3}.

The pure-gravity sector of the SME has been studied
to seek out possible experimental signals
of Lorentz violation \cite{gr4}
in addition to the ones known in flat spacetime.
Under simple assumptions
there are 18 coefficients of relevance to pure-gravity tests
that have not been tested in the existing context of the PPN formalism.
A number of tests are of interest,
including ones involving lunar and satellite laser ranging,
laboratory tests with gravimeters and torsion pendula,
measurements of the spin precession of orbiting gyroscopes,
timing studies of signals from binary pulsars,
and the classic tests involving the perihelion
precession and the time delay of light
\cite{gravtests}.

In this proceedings,
we focus on flat-spacetime tests of Lorentz symmetry
using spectroscopy of hydrogen and antihydrogen \cite{hbar,hbar2},
and also on related results using the Penning-trap system \cite{Penning:theory}.
Sharp bounds on a number of coefficients for Lorentz violation
have been produced using hydrogen
masers \cite{maser}.
Efforts to create antihydrogen atoms
for spectroscopic tests have progressed steadily
over a period of several years.
Trapped antihydrogen atoms were first produced experimentally
by the
ATHENA \cite{athena},
and the
ATRAP \cite{atrap}
collaborations,
based at CERN.
The ASACUSA collaboration, also at CERN,
has conducted successful related studies
of antiprotonic helium \cite{asacusa}.
The recently-formed ALPHA collaboration \cite{alpha}
is continuing the work of the ATHENA group.

\section{The Standard-Model Extension}
Since Lorentz violations are known to be small,
perturbation theory is well suited for the
calculation of effects.
In the SME, the violations enter in the form of a variety of coefficients,
which are different for each species of fundamental particle.
For practical purposes,
composite particles such as protons and neutrons are also treated as fundamental,
so that they too have coefficients for Lorentz violation.
The effective parameters include
$a^w_\mu$, $b^w_\mu$, $H^w_{\mu\nu}$, $c^w_{\mu\nu}$, $d^w_{\mu\nu},$
where $w$ indicates the particle species.
For example, we use $w=e$ for electrons and $w=p$
for protons.
For the antiparticles,
the coefficients differ by a sign in some cases \cite{SME:minkowski}.
Here, we consider calculations in the context of
relativistic quantum mechanics,
where the perturbative hamiltonian has the form
\bea
\hat H_{\rm pert}^w
&=&
a^w_\mu \ga^0 \ga^\mu - b^w_\mu \ga_5 \ga^0 \ga^\mu
\nonumber \\ &&
- c^w_{0 0} m \ga^0 - i (c^w_{0 j} + c^w_{j 0})D^j
+ i (c^w_{0 0} D_j - c^w_{j k} D^k) \ga^0 \ga^j
\nonumber \\
&& - d^w_{j 0} m \ga_5 \ga^j + i (d^w_{0 j} + d^w_{j 0}) D^j \ga_5
+ i (d^w_{0 0} D_j - d^w_{j k} D^k) \ga^0 \ga_5 \ga^j
\nonumber \\&&
+ \half H^w_{\mu \nu} \ga^0 \si^{\mu \nu}
\nonumber
\quad .
\label{Hint}
\eea
In this equation, the particle has mass $m$
and charge $q$,
the electromagnetic field is $A^\mu$,
and $D^\mu$ is the covariant derivative of the form
$i D_\mu = i \partial_\mu + q A_\mu$.
We focus on electrons and protons, and their antiparticles,
all of which are described by spinors.
The general form is $\ch_{n,s}^w$,
where $n$ is a composite index for the quantum numbers of the wave function,
and $s$ is the spin quantum number.
The first-order corrections to the energy levels
are calculated from
\beq
\de E_{n,s}^w = \int
\ch_{n,s}^{w \dagger} \, \hat H_{\rm pert}^{w} \,
\ch_{n,s}^w \, d^3r
\, ,
\label{delE}
\eeq
with a similar expression in the case of the antiparticles.
In the following,
we look at the perturbations of the energy levels
for the hydrogen and antihydrogen atoms,
and for electrons and positrons in Penning traps.
The expressions for the small energy shifts contain
information invaluable for the success of experimental detection
of potential symmetry violations.
This information includes the suppression level of the effects,
the dependence of the effects on the orientation of the quantization axis,
and the dependence of the effects on the state of motion of the system.

\section{Hydrogen and Antihydrogen}
Here, we summarize results of calculations made
to find the Lorentz-violating effects in hydrogen and
antihydrogen \cite{hbar}.
A first approach is to consider free atoms,
since this would minimize shifts in the spectral lines
due to electromagnetic fields.
In antihydrogen,
the contributions to the violation effects
arising from both the antiproton and the positron
must be considered,
and so the four possible spin states
expressed in the decoupled basis
are relevant.
We use quantum numbers $m_J=\pm 1/2$ and $m_I = \pm 1/2$,
where $J$ and $I$ are the positron and antiproton angular momenta, or,
in the case of hydrogen, the electron and proton angular momenta.
The resulting shifts in the energy levels for
the $n=1$ and $n=2$ levels of free hydrogen are:
\bea
\De E^{H} (m_J, m_I)
& \approx &
(a_0^e + a_0^p - c_{00}^e m_e - c_{00}^p m_p)
\cr
&&
+ (-b_3^e + d_{30}^e m_e + H_{12}^e) {m_J}/{|m_J|}
\cr
&&
+ (-b_3^p + d_{30}^p m_p + H_{12}^p) {m_I}/{|m_I|} \quad .
\label{EHJI}
\eea
These expressions show that the $1S$ to $2S$
transitions in hydrogen are not affected by the coefficients for Lorentz violation
at leading order, since in each case, the energy shifts in the two levels are identical.

The free case is of course not necessarily the most relevant,
since antihydrogen atoms need to be confined using suitable trapping fields.
We therefore calculate spectral shifts
in the presence of a uniform magnetic field,
which approximates the environment near the center of a realistic trap.
For both hydrogen and antihydrogen,
a uniform magnetic field $B$
splits the 1S and 2S levels into four hyperfine Zeeman levels,
given in order of increasing energy by $\ket{a}_n$, $\ket{b}_n$,
$\ket{c}_n$, $\ket{d}_n$,
with principal quantum number $n=1$ or $2$.
Only the low-field seeking $\ket c$ and $\ket d$ states are trapped
and so we consider transitions
involving these states.

In the case of the $1S$ to $2S$
transition,
the shifts in energies of the states
$\ket{d}_1$ and $\ket{d}_2$
are again found to be identical,
so no leading-order effect on this
particular transition occurs.
For the
1S-2S transition between the
$\ket{c}_1$ and $\ket{c}_2$ states,
an unsuppressed frequency shift is found.
It can be traced back to the difference in mixing angles
$\th_1$ and $\th_2$ for the two states:
\beq
\tan 2 \th_n \approx (51 {\rm ~mT})/n^3B
\, .
\eeq
The frequency shift is:
\beq
\De \nu_{1S-2S,c} \equiv \nu_c^H
- \nu_c^{\overline{H}} \approx - \ka (b_3^e - b_3^p)/\pi
\, ,
\eeq
where the function $\ka$
is defined by
\beq
\ka\equiv \cos 2\th_2 - \cos 2\th_1
\, .
\eeq
In theory,
the effect is maximal
at a magnetic field of about $B \simeq 0.01$ Tesla.
The subscript $3$ in $b_3^e - b_3^p$ refers to the quantization axis
of the system, defined by the direction of the magnetic field
of magnitude $B$.
We note that limitations due to the broadening of the spectral lines
are likely.

The hyperfine transitions for hydrogen and antihydrogen
are also of interest.
In the case of hydrogen,
the relevant $n=1$ leading-order shifts
are
\bea
\De E_a^H  &\simeq&
\hat\ka (b_3^e - b_3^p - d_{30}^e m_e
+ d_{30}^p m_p - H_{12}^e + H_{12}^p)
\quad ,
\nonumber\\
\De E_b^H &\simeq&
b_3^e + b_3^p - d_{30}^e m_e
- d_{30}^p m_p - H_{12}^e - H_{12}^p
\quad ,
\nonumber\\
\De E_c^H &\simeq& -\De E_a^H
\quad , \qquad
\De E_d^H \simeq - \De E_b^H
\quad ,
\label{abcd}
\eea
where
\beq
\hat\ka \equiv \cos2 \th_1
\, .
\eeq
In the case of the
$\ket{d}_1 \longrightarrow \ket{c}_1$ transition,
the magnetic-field dependence,
entering through the function $\hat\ka$,
vanishes at the value of about 0.65~Tesla.
Other techniques may also be experimentally relevant
to eliminate line broadening.
At this magnetic-field strength,
the transition is mostly a proton spin flip
and so any Lorentz-violating effect would be
predominantly due to proton coefficients.
The leading-order shifts
in the frequencies
$\nu_{c \rightarrow d}^H$ and
$\nu_{c \rightarrow d}^{\overline{H}}$
for hydrogen and antihydrogen respectively are:
\bea
\de \nu_{c \rightarrow d}^H
&\approx&
(-b_3^p + d_{30}^p m_p + H_{12}^p)/\pi
\quad , \label{freq:h:cd}\\
\de \nu_{c \rightarrow d}^{\overline{H}}
&\approx&
(b_3^p + d_{30}^p m_p + H_{12}^p)/\pi
\quad . \label{freq:Hbar:cd}
\eea

Hyperfine transitions in other systems are also of interest,
because they can be expected to have qualitatively similar effects.
The ASACUSA collaboration has made measurements of the hyperfine structure
in antiprotonic helium \cite{hfs:asacusa}.

\section{Time dependence in Lorentz-violating signals}
In expressions like \rf{freq:h:cd} and \rf{freq:Hbar:cd} above,
the subscripts on the coefficients for Lorentz violation
refer to a laboratory-fixed reference frame.
This frame is approximately inertial if the experiment is run over a few hours,
but over an extended period the rotation
and direction of motion of the laboratory has to be accounted for.
The SME coefficients for Lorentz violation
are considered to be constant
in the Sun-based standard inertial reference frame
that has been adopted for tests of the SME.
Due to the motion of the laboratory relative to this frame,
the experimental observables are found to be time dependent.
For any Earth-fixed laboratory,
this dependence includes the sidereal period
of just under 24 hours.
For satellites, the periodicities are different \cite{cc:space}.
In a number of recent experiments with microwave and optical cavities,
rotating turntables have been used to introduce other periodicities
with the intention of making the effects
more readily detectable \cite{photon:cavities}.

The use of a standard reference frame $(T,X,Y,Z)$
has made it possible to compare results from different experiments
and to organize results from a variety of different areas
in a unified manner.
The transformations relating the laboratory-frame coordinates
$(t,x,y,z)$ to the inertial one
are discussed in Appendix C of the third article in Reference
\cite{km}.

Since the signals of Lorentz violation
are small shifts in frequencies,
experiments are done by comparing one frequency with another.
One approach is to continuously make this type of comparison
and attempt to discern a periodic drift with a sidereal frequency
of just under 24 hours.
This sidereal method could be done, for example, by comparing
$\nu_{c \rightarrow d}^{\overline{H}}$, which has a leading order effect,
to $\nu_{a \rightarrow c}^{\overline{H}}$, which has no leading order effect.
Another approach can be adopted in cases where measurements
that are effectively instantaneous can be made.
Then, for example, one could compare
$\nu_{c \rightarrow d}^{\overline{H}}$ for antihydrogen
with $\nu_{c \rightarrow d}^{H}$
done with conventional
antihydrogen, where the quantization axes are identical.

Sidereal tests seeking Lorentz violation
in conventional hydrogen have been done
using a hydrogen maser \cite{maser}.
The $F=1$, $\De m_F=\pm 1$ transition
was used with a weak  magnetic field,
placing a bound
at the level of about $10^{-27}$~GeV
on a mixture of electron and proton parameters
of the form in equation \rf{freq:h:cd}.
Sidereal tests with antihydrogen
may become possible in the future.
These could in principle be done in a similar manner
by monitoring the hyperfine transition and comparing it
to a suitable stable reference frequency.
This would bound the combination of Lorentz-violation coefficients
seen in equation \rf{freq:Hbar:cd}.

Instantaneous comparisons of the hyperfine transitions
in hydrogen and antihydrogen would bound the difference between the two
quantities in
(\ref{freq:h:cd}) and (\ref{freq:Hbar:cd}):
\beq
\De \nu_{c \rightarrow d} \equiv
\nu_{c \rightarrow d}^H - \nu_{c \rightarrow d}^{\overline{H}}
\approx - 2 b_3^p / \pi
\quad .
\eeq
This type of test
places a bound on just one coefficient for Lorentz violation
and so offers some advantages over sidereal tests.
We note that the $b_3^p$ coefficient quantifies both Lorentz and CPT violation.
A bound at the level of $10^{-26}$~GeV
would be obtained if a 10-mHz resolution was achieved in the spectral line.

We note that the hyperfine transitions in the microwave regime
can attain a given bound on the $b_3$-type coefficients
with a lower fractional precision than the 1S-2S optical transitions.
The 1S-2S optical transitions, at about  $10^{15}$ Hz,
require a part in $10^{15}$ fractional precision to get a 1 Hz resolution,
whereas the singlet to triplet transition in the hyperfine structure
occurs at about $10^9$ Hz, and so a part in $10^9$
fractional precision will attain a 1 Hz resolution in this case.

\section{Penning Traps}
Many aspects of the above discussion
for hydrogen and antihydrogen
apply also for Lorentz-symmetry tests
using trapped fermions in Penning traps.
In particular,
we summarize here
some relevant details for
tests based on measurements of the anomaly frequency
in the case of trapped electrons and positrons
\cite{Penning:theory}.

Basically,
Penning traps have a strong uniform magnetic field
serving to confine charged particles
to a region close to a
central symmetry axis.
The particles, which are of one sign only,
are prevented from drifting along the $B$-field direction by
an electric field,
which in some cases is a quadrupole field.
This device
is capable of trapping a single particle for as long as
several months,
and can be used to make high-precision measurements
of oscillation frequencies of the particle.

As in the case of hydrogen and antihydrogen,
the possible effects of Lorentz violation in this system
have been
calculated using the SME framework in perturbation theory
\cite{Penning:theory}.
Since the energy-level pattern
is determined predominantly by the magnetic field,
the calculations are simplified by using the
relativistic Landau levels of the particle
in a uniform magnetic field as the unperturbed states.
One of the findings is that the cyclotron frequency
is not affected at leading order by Lorentz violation.
The most significant results from a theoretical standpoint
are the unsuppressed shifts found in the
anomaly frequencies.
For the electron,
the shift is
\beq
\om_a^{e^-} \approx \om_a
- 2 b_3^e + 2 d_{30}^e m_e + 2 H_{12}^e
\equiv \om_a - 2 \tilde b_3^e
\quad .
\label{waelec}
\eeq
The definition of $\tilde b_3^e$
is consistent with definitions used in the context of clock-comparison
experiments \cite{cc},
which are limited to experiments using matter
but not antimatter.
We note that $\tilde b_3^e$
has been finely bounded using
a torsion-pendulum experiment \cite{torsion:exp}.
For the positron,
the anomaly-frequency shift is
\beq
\om_a^{e^+} \approx \om_a
+ 2 b_3^e + 2 d_{30}^e m_e + 2 H_{12}^e
\quad .
\label{wapositron}
\eeq
This particular combination of SME coefficients
cannot be directly bounded by experiments using
torsion pendula or atomic clocks,
since it requires antimatter.
We note that the similarity
of equations \rf{waelec} and \rf{wapositron}
to equations \rf{freq:h:cd} and \rf{freq:Hbar:cd}
indicates similarities in the physics,
since both signal types are based on spin-flip transitions
in a uniform magnetic field.
Again, the subscripts on the coefficients for Lorentz violation
refer to laboratory frame coordinates,
which are not inertial unless the experiment can be completed
in no more than an hour or two.
Sidereal-type experiments
would monitor the anomaly frequency and compare it
to a reference frequency such as the cyclotron frequency.

Instantaneous-type experiments would compare the anomaly frequency
of the electron with that of the positron,
under the assumption that the magnetic field remains constant.
If this can be accomplished to a reasonable approximation experimentally,
the quantity measured would be the frequency difference:
\beq
\De \om_a^e  \equiv \om_a^{e^-} - \om_a^{e^+} \approx - 4 b_3^e
\quad .
\label{delwce}
\eeq
We note that the units employed here assume that $\hbar=1$.
As with antihydrogen and hydrogen Lorentz tests,
it is the absolute frequency resolution that
determines the sharpness of the test,
since the frequency resolution translates directly
via Planck's constant into the bound on the SME quantity.

Several experimental results have been found using Penning traps,
which attained resolutions of a few hertz
\cite{Penning:expt}.
More recent technology should be able to improve on these results.

Although antiprotons have been trapped
in Penning traps for antihydrogen experiments,
single-particle precision frequency measurements aimed at testing
Lorentz symmetry may be more difficult than for electrons and positrons.
Experiments with protons and antiprotons
are expected to improve on existing measurements of
the gyromagnetic ratios of both particles
\cite{ptexp2}.

\section{Discussion}
The SME is a broad framework for testing Lorentz symmetry in nature.
It has provided the basis for numerous experimental tests over the last decade.
The coefficient space for Lorentz violations
still has significant untested regions.
A useful feature of the SME framework is its ability to unify results
from seemingly disparate experiments.
For example, the coefficient-space region accessed in experiments with
hydrogen and antihydrogen
has an overlap with the regions tested
in experiments with Penning-trap experiments involving electrons and positrons.
Torsion-pendulum experiments have also accessed portions of this region.
Data tables for Lorentz violation can be found in Ref.\ \cite{tables}.

Calculations to discern the effects of Lorentz violation on the spectrum
of hydrogen and antihydrogen
have been performed.
One of the findings is that
suppression effects are reduced in transitions involving
spin flips.
Since these transitions are also affected by magnetic fields,
there are experimental challenges to face.
Possibilities for good Lorentz tests
include ones based on the hyperfine transitions.
In cases where the test can be done using both
hydrogen and antihydrogen,
the tests can be expected to test the discrete CPT symmetry
as well as Lorentz symmetry.

\end{document}